\documentclass[useAMS]{mn2e}
\usepackage{epsfig}
\usepackage{epsf}
\usepackage{graphicx}

\title[SNe IIn-P and the Crab Nebula]{The Crab Nebula and the class of
  Type IIn-P supernovae caused by sub-energetic electron capture
  explosions}

\author[Smith]{Nathan Smith\thanks{Email: nathans@as.arizona.edu} \\
  Steward Observatory, 933 N. Cherry Ave., Tucson, AZ 85721, USA}

\begin{document}
\date{Accepted 0000, Received 0000, in original form 0000}
\pagerange{\pageref{firstpage}--\pageref{lastpage}} \pubyear{2002}
\def\arcdeg{\degr}
\maketitle
\label{firstpage}

\begin{abstract}

  What sort of supernova (SN) gave rise to the Crab Nebula?  While
  there are several indications that the Crab arose from a
  sub-energetic explosion of an 8-10 $M_{\odot}$ progenitor star, this
  would appear to conflict with the high luminosity indicated by
  historical observations.  This paper shows that several well-known
  observed properties of the Crab and SN~1054 are well-matched by a
  particular breed of Type~IIn supernova.  The Crab's properties are
  best suited to the Type IIn-P subclass (Type~IIn spectra with
  plateau light curves), exemplified by SNe~1994W, 2009kn, and 2011ht.
  These events probably arise from relatively low-energy (10$^{50}$
  erg) explosions with low $^{56}$Ni yield that may result from
  electron-capture SN (ecSN) explosions, but their high
  visual-wavelength luminosity and Type~IIn spectra are dominated by
  shock interaction with dense circumstellar material (CSM) rather
  than the usual recombination photosphere.  In this interaction, a
  large fraction of the 10$^{50}$ ergs of total kinetic energy can be
  converted to visual-wavelength luminosity.  After about 120 days,
  nearly all of the mass outside the neutron star in the CSM and
  ejecta ends up in a slowly expanding (1000-1500 km s$^{-1}$) thin
  dense shell, which is then accelerated and fragmented by the growing
  pulsar wind nebula (PWN) in the subsequent 1000 yr, producing the
  complex network of filaments seen today.  There is no need to invoke
  the extended, invisible fast SN envelope hypothesized to reside
  outside the Crab.  As differentiated from a normal SN~II-P,
  SNe~IIn-P provide a much better explanation for several observed
  features of the Crab: (1) No blast wave outside the Crab Nebula
  filaments, (2) no rapidly expanding SN envelope outside the
  filaments, (3) a total mass of $\sim$5 $M_{\odot}$ swept up in a
  thin slow shell, (4) a low kinetic energy of the Crab at least an
  order of magnitude below a normal core-collapse SN, (5) a high peak
  luminosity ($-$18 mag) despite the low kinetic energy, (6) chemical
  abundances consistent with an 8-10 $M_{\odot}$ star, and (7) a low
  $^{56}$Ni yield.  A number of other implications are discussed,
  concerning other Crab-like remnants, the origin of dust in the Crab
  filaments, diversity in the initial masses of SNe~IIn, and the
  putative association between ecSNe and SN impostors.  This model
  predicts that if/when light echoes from SN~1054 are discovered, they
  will exhibit a Type IIn spectrum, probably similar to SNe~1994W and
  2011ht.

\end{abstract}


\section{INTRODUCTION}

The Crab pulsar proves that SN~1054 must have marked the final
core-collapse supernova (SN) explosion of a massive star --- but what
exactly was the initial mass of that star, and what were the
properties of the explosion that gave rise to the Crab Nebula we see
today?  Answers to these basic questions remain highly uncertain,
despite decades of intensive and careful observations.  A central
mystery that has never been resolved is that while SN~1054 was more
luminous than a normal Type II SN, the kinetic energy of the Crab
Nebula is surprisingly low (about 7$\times$10$^{49}$ erg or less)
compared to the canonical 10$^{51}$ ergs of kinetic energy in a
typical SN.

The standard explanation for this fundamental puzzle of the Crab,
summarized recently by Hester (2008), is that most of the mass and
90\% of the kinetic energy of SN~1054 actually reside far {\it
  outside} the visual nebulosity known as the Crab Nebula, in an
invisible freely expanding envelope of cold and neutral SN ejecta.
This fast envelope, as well as the blast wave one expects at the
leading edge of the fast ejecta that collide with the ambient medium,
have never been detected to remarkably low upper limits (see below).
In this scenario, first articulated by Chevalier (1977), the Crab
Nebula that we see is only the thin interface between the expanding
synchrotron nebula and the ejected stellar envelope.  Some models for
the Crab involve an ``electron-capture SN'' (ecSN) marking the
collapse of a degenerate ONeMg core in a star with initial mass 8-10
$M_{\odot}$ (e.g., Nomoto et al.\ 1982; Nomoto 1987; Miyaji et al.\
1980)\footnote{The range of initial mass for ecSNe varies between
  studies; some prefer masses near the higher end of this range (see,
  e.g., Wanajo et al.\ 2009).}, producing a weaker explosion
(typically 10$^{50}$ erg of kinetic energy, instead of 10$^{51}$ for
Fe core collapse).  These models, however, also predict sub-luminous
explosions (Kitaura et al.\ 2006) due to the under-production of
$^{56}$Ni compared to Fe core-collapse SNe.  Thus, the disagreement
between the Crab's low kinetic energy and the high luminosity of
SN~1054 has remained quite puzzling.

This paper proposes a solution to this enduring mystery, enlisting
shock interaction with dense circumstellar material (CSM) as seen in
many Type~IIn SNe.  CSM interaction can produce a very luminous SN
despite a low total explosion kinetic energy, because a high fraction
(typically $\sim$10-30\% or more) of the kinetic energy is converted
to radiation. This same model matches many other unusual properties of
the Crab, like the slow, thin, filamentary shell that is the direct
product of dense CSM interaction, and helps to provide a unifying
picture of SN~1054.

The Crab Nebula is one of the most carefully observed objects in the
sky, providing a long list of detailed observational constraints for
any model.\footnote{Here we concentrate on the thermal filamentary
  shell of the Crab Nebula in order to diagnose properties of SN~1054.
  We do not discuss the remarkable properties of the synchrotron
  nebula or the central pulsar, except with respect to their role in
  shaping and accelerating the filaments long after the SN.}  An
abbreviated list of some of the most relevant observed properties of
SN~1054 and the Crab Nebula are given below.  These are compiled from
two comprehensive reviews by Davidson \& Fesen (1985) and Hester
(2008), except where noted.

1. SN~1054 was certainly a core-collapse event because it produced a
neutron star.  This could, however, be accomplished either by a
standard Fe core collapse, or by an ecSN event.  There are a number of
reasons to favor the latter.

2.  SN~1054 should have been a Type II supernova (had spectra been
available), because H emission lines are bright in the Crab filaments.
The relative n(H)/n(He) abundance of $\sim$2 (Davidson \& Fesen 1985)
is too high for it to have been a Type~IIb event, and certainly not a
Type~Ib or Ic (Dessart et al.\ 2012; Haschinger et al.\ 2012).  The
only viable known types are then Type II-P, II-L, or IIn.

3.  SN~1054 was, however, quite luminous compared to normal SNe~II-P
and IIb, with a peak absolute visual magnitude of roughly $-$18 (this
is discussed in more detail below in \S 2).  The average peak
luminosity for SNe II-P is around $-$15.6, and around $-$16.7 for SNe
IIb (Li et al.\ 2011).

4.  The network of thermally emitting filaments that encloses the
synchrotron nebula has an expansion rate suggestive of an origin in
SN~1054.  While proper motions of the filaments indicate a later
ejection date of 1120-1233 A.D.\ when extrapolated back in time
assuming linear expansion (Trimble 1968; Wyckoff \& Murray 1977;
Bietenholz et al.\ 1991; Nugent 1998), it is thought that the pulsar
wind nebula (PWN) has been pushing outward against the filaments and
has accelerated them (e.g., Woltjer 1958), producing a younger
apparent age and driving the instabilities that shape the filaments.
More recently, Rudi, Fesen, \& Yamada (2007) have measured the proper
motion of the Crab's northern ``jet'', which is more distant from the
pulsar and less likely to be influenced by acceleration from the PWN,
and they measure an ejection date of 1055~A.D.\ $\pm$24 yr.

5.  The total mass in the Crab Nebula's filaments is difficult to
estimate from nebular spectroscopy, but is likely to be around 5
$M_{\odot}$, although some estimates have been lower (1--2
$M_{\odot}$).  Some additional mass might be hidden in the dense
neutral/molecular cores of some filaments that are shielded from UV
radiation and not traced by the emission lines used to estimate the
mass.  Indeed, high resolution images of filaments and blobs do show
ionization gradients consistent with self-shielding (e.g., Blair et
al.\ 1997; Sankrit et al.\ 1998).  We adopt $\sim$5 $M_{\odot}$ as the
total mass in the shell of filaments.

6.  The Crab Nebula is expanding slowly, and has extremely low kinetic
energy for a SN.  Drift scans of the integrated spectrum from the
entire nebula show that most of the gas producing the brightest
visual-wavelength emission lines is expanding at roughly $\pm$1200 km
s$^{-1}$ (MacAlpine et al.\ 1989; Fesen et al.\ 1997; Smith 2003),
which we adopt as the representative bulk speed for most of the mass
in the filaments.  Fainter emission extends to $\pm$2000-2500 km
s$^{-1}$, but only very low-level emission and a small fraction of the
mass extends to that speed.  With $\sim$5 $M_{\odot}$ expanding at
roughly 1200 km s$^{-1}$, the kinetic energy of the Crab filaments is
only 7$\times$10$^{49}$ ergs (and this is more generous than some
estimates).

7.  No blast wave has ever been detected outside the Crab Nebula in
X-ray emission (Mauche \& Gorenstein 1989; Predehl \& Schmitt 1995;
Seward, Gorenstein, \& Smith 2006) or radio emission (Frail et al.\
1995).  A blast wave located outside the Crab at many times the radius
of the current filaments is expected in the ``standard'' model for the
Crab Nebula, where a freely expanding, fast, and cold stellar envelope
should be colliding with the ambient medium.  Attempts to explain the
lack of a blast wave require special conditions in the surrounding
medium and are generally unsatisfying.

8.  No neutral envelope outside the visible filaments with speeds
comparable to a normal SN~II-P has ever been detected.  As above, an
extended fast expanding envelope of cold SN ejecta should reside
outside the Crab in the standard view.  Lundqvist et al.\ (1986)
predicted that because of ionization from the synchrotron emission,
this extended envelope would be easily detected in a number of UV
absorption lines. Sollerman et al.\ (2000) presented a weak detection
of the C~{\sc iv} $\lambda$1550 resonance lines in absorption, but did
not detect any material moving faster than 2500 km s$^{-1}$, only
tracing about 0.3 $M_{\odot}$ of material at those relatively slow
speeds that correspond to the outer parts of the visible filaments.
Other upper limits on any fast envelope outside the Crab are quite
restrictive (e.g., Fesen et al.\ 1997), making it unlikely that there
is an envelope of several $M_{\odot}$ moving at speeds up to 10,000 km
s$^{-1}$ as one expects if SN~1054 were a normal SN II-P.

9.  Analysis of the chemical composition of the Crab Nebula filaments
reveals them to be He and C rich, and not as O-enriched as other ccSN
remnants (see Davidson \& Fesen 1985, and references therein; e.g.,
Henry \& MacAlpine 1982; Pequignot \& Dennefeld 1983; Davidson et al.\
1982; Fesen \& Kirshner 1982; Kirshner 1974; Satterfield et al.\
2012). This yield implies a relatively low initial mass for the
progenitor star of around $M_{ZAMS} \simeq$ 8$-$10 $M_{\odot}$ (some
put the initial mass at the higher end of this range; e.g., MacAlpine
\& Satterfield 2008).

10.  The composition of the filaments also indicates relatively low
abundances of iron-peak elements, which in turn implies a low yield of
$^{56}$Ni.  This too is consistent with a relatively low-energy
ecSN.\footnote{Wanajo et al. (2009) point out that the apparent high
  Ni/Fe ratio inferred for the Crab filaments would also be consistent
  with the ecSN model.}

11.  The Crab is located about 180 pc from the Galactic plane.  Aside
from the remote possibility that it could be a runaway from the I
Geminorum association (although radial velocities seem to discount
this; Minkowski 1970), it does not appear to be associated with any
group of OB stars.  These facts also seem to point toward a relatively
low-mass progenitor star.

12. The Crab filaments formed significant amounts of both dust and
molecules, indicating that they went through a phase with very high
density and rapid cooling.  Dust in the filaments can be inferred
based on the absorption it produces, causing some of the filaments to
be seen in silhouette against the synchrotron nebula (Fesen \& Blair
1990). Far-infrared (IR) thermal emission from dust was recognized
early-on (Glaccum et al.\ 1982; Marsden et al.\ 1984).  The dust mass
has been constrained by more recent {\it Spitzer} and {\it Herschel}
observations (Temim et al.\ 2006, 2012; Gomez et al.\ 2012), with a
surprisingly large mass of 0.1--0.2 $M_{\odot}$ indicated by the {\it
  Herschel} data (Gomez et al.\ 2012).\footnote{The dust mass in the
  Crab has sparked some recent controversy. While Temim \& Dwek (2013)
  suggest that the large excess far-IR flux detected by {\it Hershel}
  does not indicate a large dust mass, P.\ Owens (2013, private comm.)
  performed an independent model analysis and finds a large dust mass
  consistent with the analysis by Gomez et al.\ (2012).} IR
spectroscopy also reveals the presence of H$_2$ in the Crab filaments
(Graham et al.\ 1990; Loh, Baldwin, \& Ferland 2010; Richardson et
al.\ 2012).

Altogether, the observed parameters listed above present a serious
challenge to understanding SN~1054 as a normal Type II explosion
(either SN~II-P, IIb, or II-L).  A large number of unrelated and
unlikely circumstances would need to conspire to produce the Crab
Nebula from a normal core-collapse SN.

In stark contrast, all of the observational properties listed above
are expected and well matched by a particular observed class of SN -
namely, the sub-class of Type~IIn-P explosions (Type~IIn spectra with
plateau light curves), discussed recently by Mauerhan et al.\ (2013a).
SNe~IIn are fundamentally different from other types of SNe because
they are dominated by intense CSM interaction, which sweeps up most of
the mass into a cold, dense shell (CDS) that collapses as a result of
radiative cooling.  It is precisely this radiative cooling that can
make SNe IIn very luminous, even with low kinetic energy.  For
SNe~IIn-P, evidence suggests that at least some members of this class
are sub-energetic (perhaps even being associated with ecSNe, although
this remains unproven), while still being as luminous as normal SNe
(or moreso).  Details of this comparison are outlined below.  We begin
with a discussion of the historical light curve and a comparison to
modern SNe (\S 2), followed by a simplified model and sequence of
events that accounts for the observed properties (\S 3).  This is
followed by a discussion of a number of further implications for the
connection between the Crab and SNe IIn-P, including a prediction for
the observed spectrum if light echoes are detected from SN~1054, as
well as related implications for the interpretation of SNe~IIn and SN
impostors in general.

\begin{figure}\begin{center}
   \includegraphics[width=3.1in]{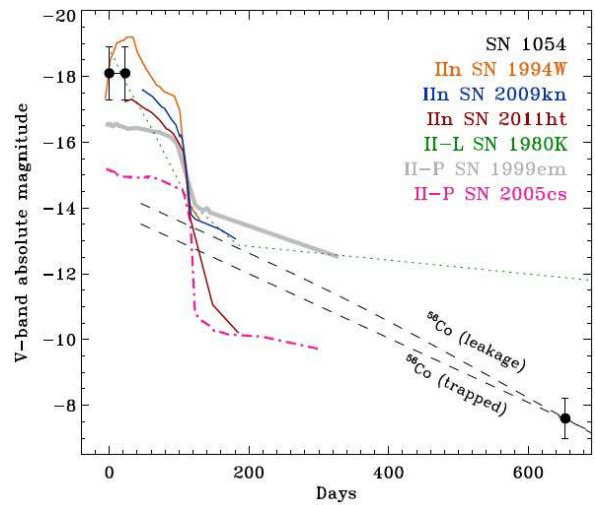}
\end{center}
\caption{The $V$-band absolute-magnitude light curves for a number of
  modern SNe compared to the limited historical information about the
  visual-wavelength luminosity of SN~1054 (black dots; see text).
  SN~1994W is from Sollerman et al.\ (2000), SN~2009kn is from Kankare
  et al.\ (2012), and SN~2011ht is from Mauerhan et al.\ (2013a).
  SN~1980K is from Buta (1982), with very late-time $V$-band
  photometry from Sugerman et al.\ (2012).  SN~1999em is from Leonard
  et al.\ (2002), and SN~2005cs is from Pastorello et al.\ (2009).}
\label{fig:lc}
\end{figure}
\section{THE HISTORICAL LIGHT CURVE}

The association of the Crab Nebula with the famous ``Guest Star''
observed by Chinese astrologers, discovered on 1054 July 4, has been
recounted many times (Lundmark 1921; Duyvendak 1942; Mayall \& Oort
1942; Shklovsky 1968; Minkowski 1971; Clark \& Stephenson 1977;
Chevalier 1977; Brecher et al.\ 1983).  The weirdness of the Crab
compared to other traditional SNe and SN remnants has also been
repeated many times, by these same authors and others. Adopting the
known extinction of $A_V = 1.6$ mag (Miller 1973) and a distance of
$\sim$2 kpc (Trimble 1973; or 1500--2200 pc as given by Davidson \&
Fesen 1985), one can translate the historical accounts of Chinese
astrologers to approximate absolute magnitudes (roughly $V$-band).
There are essentially two facts that we gain from the historical
reports:

\begin{itemize}

\item {\it SN~1054 was visible during daytime for a total of 23 days.}
  Following Shklovsky (1968), this implies $m_{vis} \simeq m_V \simeq
  -5$ mag, and therefore an absolute visual magnitude of about $-$18.1
  or brighter during those 23 days.  It is not known how day 23
  compares to the time after explosion, because the discovery on July
  4 may have been the date when SN~1054 first appeared in the early
  morning after being behind the Sun (it is a fall and winter object).
  Thus, the peak absolute magnitude is not firmly established (i.e.,
  it is at least as luminous as $-$18 mag during that time, but could
  have been more luminous at earlier times).  The time when SN~1054
  was seen during the day is marked by the first two black dots in
  Figure~\ref{fig:lc}, adopting an uncertainty of $\pm$0.8 mag
  (although the first point should perhaps be regarded as a lower
  limit to the initial luminosity).

\item {\it SN~1054 faded thereafter, and remained visible at night to
    the unaided eye until 653 days after discovery.}  The rate of
  fading is not known, but we can conclude that by day 653 the SN had
  faded to a luminosity of about $M_V$ = $-$7.6 mag.  This is
  indicated by the last black dot on Figure~\ref{fig:lc}, adopting a
  somewhat smaller uncertainty of $\pm$0.6 mag).

\end{itemize}

What do these two facts tell us about the type of SN that made the
Crab?  Before considering that question, let's remember that we can
rule out SNe~Ib and Ic, and probably also IIb, because of the presence
of substantial amounts of H in the Crab filaments.  Considering only
SNe of Type II, then, Figure~\ref{fig:lc} does provide some useful
constraints.

Figure~\ref{fig:lc} allows us to rule out normal SNe~II-P like
SN~1999em and fainter SNe~II-P like SN~2005cs because they never
achieve such a high peak luminosity.  This is quite useful
information, because one class of SNe that is consistent with {\it
  some} available information about the Crab is the group of
low-luminosity SNe II-P, which could be caused by low-energy ecSNe
that would agree with the kinetic energy and abundances of the Crab,
and its implied low-mass progenitor star.  However, the high peak
luminosity of $-$18 mag clearly rules these out (at least without some
mechanism to make them more luminous).

There are some relatively luminous SNe II-P (Li et al.\ 2011;
Elias-Rosa et al.\ 2009; Arcavi et al.\ 2012) that approach the peak
luminosity of SN~1054, but these are thought to be luminous because of
an energetic explosion that synthesizes a relatively large mass of
$^{56}$Ni.  A large mass of $^{56}$Ni would produce a luminous
radioactive decay tail that would be too bright (in fact, according to
Figure~\ref{fig:lc} even SN~1999em would be too luminous at late
times).  Moreover, a large mass of $^{56}$Ni is in conflict with the
observed abundances of the Crab, which are deficient in Fe-group
elements (see above).  This means that attributing the late-time
faintness to dust formation alone does not solve the problem.  Some
SNe~II-L do have brighter peak luminosities around $-$17 to $-$18 mag,
but they tend to fade slowly and are more luminous at late times than
SN~1054.  The canonical SN~II-L 1980K is shown in Figure~\ref{fig:lc}
for comparison.  Normal SNe IIn like SN~1998S have light curves very
similar to SNe II-L, and do achieve peak luminosities like SN~1054.
However, these also tend to fade more slowly, and remain luminous at
late times due to ongoing CSM interaction.

Finally, there is an observed class of SNe that does seem to agree
well with the historical light curve of SN~1054, and that is the class
of SNe~IIn-P (Mauerhan et a.\ 2013a).  The $V$-band light curves of
SN~1994W, SN~2009kn, and SN~2011ht are shown in Figure~\ref{fig:lc}.
Like traditional SNe~IIn, they can achieve relatively high peak
luminosities in agreement with SN~1054, and indeed they often fade
significantly during their ``plateau''.  What is unique about these
SNe is that, unlike other SNe~IIn, their luminosity plummets after
$\sim$120 days so that they do not maintain the high level of CSM
interaction luminosity at late times seen in most SNe~IIn.  It has
been suggested (Mauerhan et al.\ 2013a; Sollerman et al.\ 2001) that
this rapid fading may indicate unusually low yields of $^{56}$Ni, or
efficient dust formation (or both).  This drop should also be
accompanied by a drop in CSM interaction, although the reason why all
these SNe would do this at $\sim$120 days is not obvious (Mauerhan et
al.\ 2013a).  The drop from the end of the plateau is more extreme in
some SNe~IIn-P as compared to others.  The most extreme drop after the
plateau is for the case of SN~2011ht, but in this case we know that
the colors became very red, whereas the bolometric luminosity did not
drop quite so much (Mauerhan et al.\ 2013a).  Once again, dust
formation may be influential, whereas SN~2009kn may have formed less
dust and consequently had bluer colors on its decay tail.  The rapid
drop after the plateau probably correlates with low $^{56}$Ni mass, as
in low-luminosity SNe~II-P.

For comparison, Figure~\ref{fig:lc} also includes representative
fading rates due to radioactive decay at late times (black dashed
lines), matched to the one late-time luminosity constraint for
SN~1054.  Two cases are shown: (1) all the $\gamma$-ray luminosity is
trapped by the ejecta and reprocessed into visual light, which would
indicate a mass of synthesized $^{56}$Ni of about 0.009 $M_{\odot}$,
and (2) some leakage of the $\gamma$-rays produced by radioactive
decay (where the $\gamma$-ray optical depth evolves as $\tau \propto
t^{-2}$, following Sollerman et al.\ 2001), yielding a somewhat larger
mass of $M$($^{56}$Ni) = 0.016 $M_{\odot}$.  Both are significantly
less than the $^{56}$Ni mass expected from a normal 10$^{51}$ erg SN
event.  As we discuss below, however (and as mentioned by Sollerman et
al. 2001), CSM interaction may also make a contribution to the late
time luminosity, potentially lowering the $^{56}$Ni mass further.

How does the apparent uniqueness of the Crab compare to the frequencey
of various SN subtypes?  The Crab Nebula is the only one of its kind
among the handful of known Galactic SN remnants that are connected to
historically observed events, provoking many commentaries about how
unique or bizarre it is.  Considering the statistics, we note that SNe
of Type~IIn make up only 8-9\% of all core-collapse SNe in a
volume-limited sample (Smith et al.\ 2011a), and an even lower
fraction of all SNe if Type Ia events are included (Li et al.\ 2011).
Moreover, the SN~IIn-P sublass represents only a subset of all
SNe~IIn.  The apparently unique character of the Crab among young
nearby SN remnants in the Galaxy is therefore not so surprising if
SN~1054 was a Type~IIn-P event, but it is very difficult to reconcile
with the idea that SN~1054 was a normal SN~II-P in which most of the
mass and kinetic energy have gone undetected at the present time.
This is diuscussed further in \S 4.2 and \S 4.5.

\section{CSM  INTERACTION LUMINOSITY}

The observational case outlined in the previous section makes a
compelling argument that SN~1054 shared properties in common with
SNe~IIn, and the sub-class of SNe~IIn-P in particular.  We must
discuss whether this makes sense physically as well as
observationally.  This section outlines a very simple baseline model
of CSM interaction that can account for both the luminosity of the
events (SN~1054 and the class of SNe~IIn-P) and the resulting mass,
kinematics, and structure of the Crab Nebula.

\subsection{A Simple Model for SNe IIn-P}

Let's begin with some order-of-magnitude estimates.  For SNe~IIn-P and
for SN~1054, we wish for the shock interaction between the explosion
ejecta and the dense CSM to achieve a luminosity of order $-$18 mag or
$\sim$10$^{9}$ $L_{\odot}$ for the first few months.

In this scenario, dense CSM decelerates the shock, and the resulting
high densities in the post-shock region allow the shock to become
radiative (i.e. the shock becomes momentum conserving instead of
energy conserving, as in all SNe~IIn).  With high densities and
optical depths in H-rich gas, thermal energy is radiated away
primarily as visual-wavelength continuum emission.  This loss of
energy removes pressure support behind the forward shock, leading to a
very thin, dense, and rapidly cooling shell at the contact
discontinuity (usually referred to as the ``cold dense shell'', or
CDS; see Chugai et al.\ 2004; Chugai \& Danziger 1994).  This CDS is
pushed by ejecta entering the reverse shock, and it is slowed and
mass-loaded by the CSM, into which it expands at a speed $V_{CDS}$.
In this scenario, the maximum emergent continuum luminosity from CSM
interaction is given by

\begin{equation}
L_{CSM} \ = \ \frac{1}{2} \, \dot{M} \, \frac{V_{CDS}^3}{V_W} \ = \ \frac{1}{2} \, w \, V_{CDS}^3
\end{equation}

\noindent where $V_{CDS}$ is the outward expansion speed of the CDS,
$V_W$ is the speed of the pre-shock wind, $\dot{M}$ is the mass-loss
rate of the wind, and $w$ = $\dot{M}/V_W$ is the so-called wind
density parameter (see Chugai et al.\ 2004; Chugai \& Danziger 1994;
Smith et al.\ 2010).\footnote{Note that this scenario where radiation
  escapes efficiently is somewhat different from a more extreme case
  where the radiation diffusion time is comparable to the expansion
  timescale, changing the shape of the light curve (Smith \& McCray
  2007; Chevalier \& Irwin 2011; Falk \& Arnett 1977).}

For a radiated luminosity of 10$^{9}$ $L_{\odot}$ and for an assumed
expansion speed of the CDS of around 1100 km s$^{-1}$ (recall that the
present speed of $\sim$1200 km s$^{-1}$ in the Crab filaments is the
result of later acceleration by the PWN), we require a CSM density of
order $w$ $\approx$ 6$\times$10$^{18}$ g cm$^{-1}$.  The total mass in
the CSM would be $\sim$3 $M_{\odot}$ $R_{15}^{-1}$, where $R_{15}$ is
the outer radius of the CSM shell in units of 10$^{15}$ cm.

Note that according to Equation (1) we could also achieve a very high
CSM interaction luminosity with less dense CSM but with a faster shock
speed, due to the $v^3$ dependence.  For example, we could also reach
$L \simeq 10^9 L_{\odot}$ with $V_{shock}=10^4$ km s$^{-1}$ and
$w$=8$\times$10$^{15}$ g cm$^{-3}$.  This is, however, infeasible for
the Crab because CSM interaction must also conserve momentum, and in
this scenario with a faster shock, the lower masses of the CSM and SN
ejecta (which together would need to be less than 1 $M_{\odot}$) are
ruled out by the observed mass and speed of the Crab filaments.  A low
CDS shock speed and a low energy are needed to match the observed
constraints of the Crab, whereas we do not have independent
constraints on the CDS mass in the case of extragalactic SNe~IIn-P.

The observed visual CSM-interaction luminosity should be close to the
total bolometric luminosity during the bright ``plateau'' phase of the
event (i.e. a small bolometric correction), judging from the apparent
temperatures of $\sim$6000-7000 K typically seen in the continua of
virtually all SNe~IIn.  This may not be strictly true if the apparent
temperature changes with time as the shock slows down.  In that case
the estimated mass-loss rate or the ejecta speed might need to be
increased slightly, which doesn't significantly change the nature of
the event discussed here.  To be sure, CSM interaction does afford
some flexibility in model parameters.

One can also adjust the geometry so that the interaction has a
different strength at various latitudes.  Small differences of this
sort may account for the variation from one SN~IIn-P to the next, and
there does appear to be some departure from spherical geometry in the
CSM interaction that yielded the Crab Nebula (see below).  The purpose
here is to be illustrative rather than exact, in order to demonstrate
that CSM interaction can naturally fix the long-standing puzzle of the
high peak luminosity of SN~1054 being apparently at odds with the low
kinetic energy of the Crab.

\subsection{Origin of the Dense CSM}

Equation (1) suggests that SN~1054 requires a wind density parameter
of $w$ $\approx$ 6$\times$10$^{18}$ g cm$^{-1}$, which implies a total
CSM mass of about 3 $M_{\odot}$ within 10$^{15}$ cm of the progenitor.
Since the progenitor wind speed ($V_W$) is not known, there is
uncertainty about the mass-loss rate, the nature of the pre-SN mass
loss, and the physical state of the progenitor star.

Dense and massive CSM could arise from a sudden explosive or eruptive
pre-SN ejection event ($\sim$10$^{49}$ erg) occuring a few years
before the SN, as has been hypothesized for more luminous SNe~IIn such
as SN~2006gy (Smith \& McCray 2007, Smith et al.\ 2010a), for SN~1994W
(Chugai et al.\ 2004), and a number of other SNe~IIn. There is now
observed precedent for this type of event associated with SNe~IIn from
the documented series of pre-SN luminous blue variable (LBV)-like
events that were actually detected before the SN in the case of
SN~2009ip (see Mauerhan et al.\ 2013b and references therein).  There
was also a brief pre-SN outburst observed 2 yr before SN~2006jc
(Pastorello et al.\ 2007), but that event was a Type Ibn instead of a
Type IIn.  The fact that this type of precursor burst has been
detected in both SNe IIn and Ibn is very interesting with regard to
the Crab, however, given the high He abundance in the Crab filaments.
There is also some theoretical motivation for this type of pre-SN
eruption in stars of 8-10 $M_{\odot}$ due to Ne flashes in the
degenerate core (Chugai et al.\ 2004; Arnet 1974; Hillebrandt 1982;
Weaver \& Woosley 1979).  Alternatively, some evolution models for
super-AGB stars in this mass range become unstable due to strong He
flashes that cause the star's interior to exceed the classical
Eddington limit, possibly causing hydrodynamic ejection (Wood \&
Faulkner 1986; Lau et al.\ 2012).  Although Lau et al.\ (2012) also
note that the ensuing mass loss in this scenario could lead the star
to avoid the ecSN fate.

A second possibility is that the dense CSM would come from a steady,
short-duration wind.  For a fiducial pre-shock CSM speed of 10 km
s$^{-1}$, this would translate to a progenitor mass-loss rate of
$\sim$10$^{-1}$ $M_{\odot}$ yr$^{-1}$ blowing for $\sim$30 yr before
the SN.  This timescale would achieve the desired outer CSM radius of
$\sim$10$^{15}$ cm.  Such a high wind mass-loss rate from a relatively
low-mass star seems difficult to achieve, however, since stars of this
mass are not expected to undergo prolonged super-Eddington phases.

\begin{figure*}\begin{center}
   \includegraphics[width=4.2in]{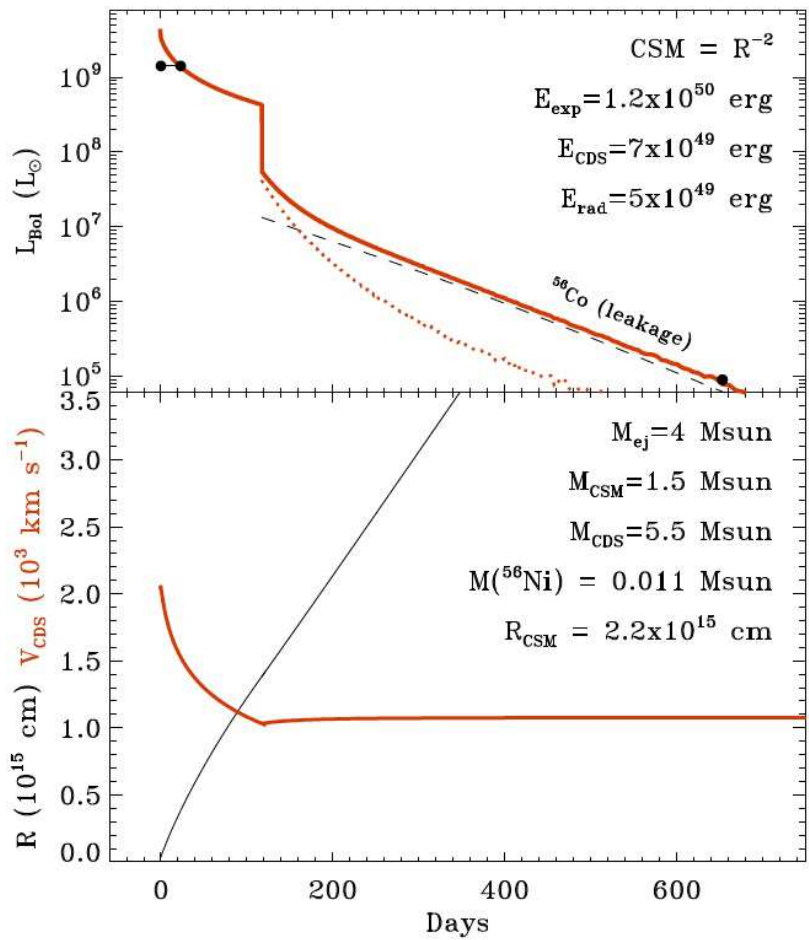}
\end{center}
\caption{Plots of a CSM-interaction model adopting a CSM with an
  $R^{-2}$ density law.  The top panel shows the radiated bolometric
  luminosity (orange), which is expected to be similar to the visual
  luminosity for a SN IIn, compared to the observational constraints
  for SN~1054 (black dots).  The solid orange line shows the total
  CSM-interaction luminosity added to the radioactive decay luminosity
  (with some leakage, same as in Figure~\ref{fig:lc}), the dashed
  black line is the radioactive decay luminosity alone, and the dotted
  orange line is the late-time luminosity from CSM interaction alone,
  excluding radioactive decay.  The bottom panel shows corresponding
  plots of the radius (black) and velocity (orange) of the CDS with
  time.  Assumed (explosion energy, ejecta mass, CSM mass, final mass
  swept into the CDS) and derived (final kinetic energy of the CDS,
  total radiated energy, synthesized $^{56}$Ni mass, outer radius of
  the CSM) physical parameters in the model are listed along the right
  side of the figure.}
\label{fig:model1}
\end{figure*}

One last alternative has properties that fit well with some
theoretical predictions for stars in the right range of initial
masses.  The CSM mass required to power SN~1054 and SNe IIn-P through
interaction might come from the dense wind of a super-AGB star with
$\dot{M} \simeq 10^{-4} M_{\odot}$ yr$^{-1}$, blowing for a few 10$^4$
yr to achieve the desired total CSM mass of $\sim$3 $M_{\odot}$.  A
few 10$^{4}$ yr duration and $\dot{M} \simeq 10^{-4} M_{\odot}$
yr$^{-1}$ is expected for the thermal-pulsing asymptotic giant branch
(TP-AGB) phase associated with super-AGB stars of initial mass 8-10
$M_{\odot}$.  In fact, precisely this scenario is found in some
stellar evolution models for the super-AGB progenitors of ecSNe (see,
e.g., Figure 15 of Poelarends et al.\ 2008).  The difficulty with this
scenario is that the massive CSM must be confined to within a few
10$^{15}$ cm in order to provide a high-enough density to power the
early light curve and the right duration of the plateau, whereas a
super-AGB wind in free expansion blowing for a few 10$^4$ yr could
extend to $\sim$0.1 pc.

Thus, the dense CSM required would seem to be either a short-duration
enhanced mass loss phase before core-collapse, or a longer-duration
wind that is somehow confined within a smaller volume (either by
external pressure of a previous hot-wind phase or gravity of a
companion star, for example).  These are unsolved issues, but it is
nevertheless true that the observed light curves of SNe IIn-P require
very dense CSM with this very compact radial extent.  Below we explore
models using CSM environments with density that falls off as $R^{-2}$
(as in a wind) and a CSM with constant density (as in a
pressure-confined wind or perhaps an ejected shell).  This distinction
turns out to cause only minor differences in the mass and energy
requirements.

\begin{figure*}\begin{center}
   \includegraphics[width=4.2in]{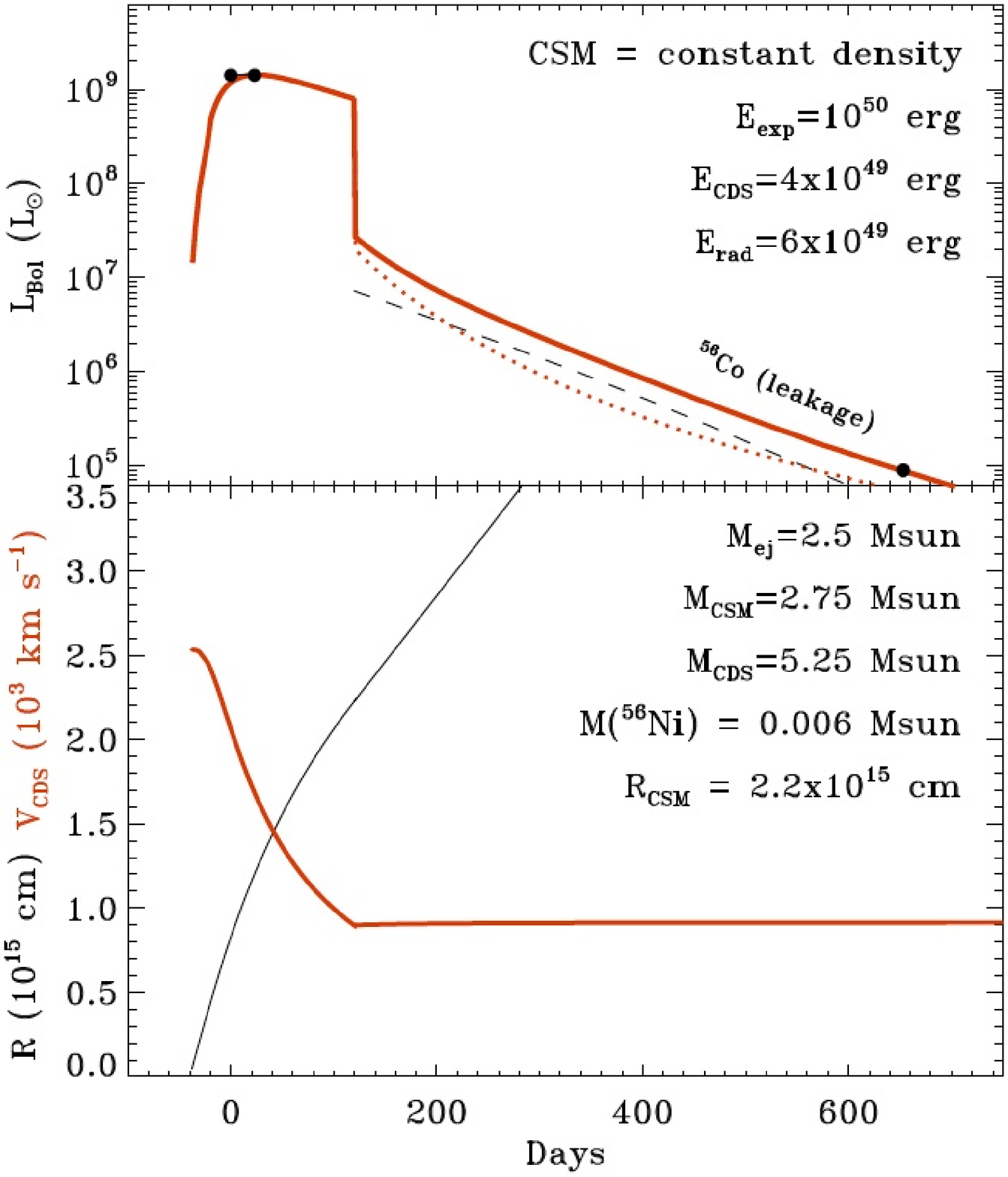}
\end{center}
\caption{Same as Figure~\ref{fig:model1}, but with a CSM that has
  constant density.  One difference is that the model has been shifted
  by $-$40 days due to the fact that the constant density CSM model
  takes a longer time for the light curve to rise to peak, and would
  likely be discovered later, relative to the $R^{-2}$ CSM model in
  Figure~\ref{fig:model1}.}
\label{fig:model2}
\end{figure*}

\subsection{The Light Curve from CSM Interaction}

Equation (1) provides a very rough estimate of the CSM density needed
to power the peak luminosity with CSM interaction.  To compute a
simple model light curve and to derive better estimates of the energy
and mass involved, however, we must account for the fact that the
speed of the CDS and the speed of SN ejecta crashing into the CDS will
change with time.  We adopt a simple spherically symmetric model where
relatively low-energy ($\sim$10$^{50}$ erg) ejecta collide with dense
CSM.  At each time-step in the ensuing collision, the CDS conserves
momentum contributed by the SN ejecta and CSM.  The deceleration of
the faster SN ejecta upon joining the CDS leads to a drop in the
kinetic energy, and the difference between the initial and final
kinetic energy at each time step provides the emergent luminosity at
that time, from which the light curve is calculated.  (Although this
light curve represents the maximum possible bolometric luminosity, we
assume that this is close to the visual luminosity since the CDS is
optically thick and its peak flux is in optical light. This should be
true during the bright plateau phase, but may become a worse
approximation at late times as the optical depth drops.) This is
essentially the same type of model used to calculate the CSM
interaction luminosity of the $\sim$10$^{50}$ erg Great Eruption of
$\eta$~Carinae (Smith 2013), except that the CSM and ejecta speeds and
mass were different in that model, which led to a fainter luminosity
with a longer duration.

We consider two cases for the CSM, with a model that has $\rho \propto
R^{-2}$, as would be appropriate for a steady short-duration wind, and
with a model that adopts a constant-density CSM envelope (as used by
Chugai et al.\ 2004 in the case of SN~1994W), which would be more
appropriate for a pressure-confined CSM or perhaps for an eruptive
event.  The resulting light curves for the two cases are shown in the
top panels of Figures~\ref{fig:model1} and \ref{fig:model2},
respectively.  The bottom panels of both figures plot the time
evolution of the radius and speed of the CDS corresponding to each
model.

The main difference between the two model light curves is in the shape
of the plateau.  Since more CSM mass is located at larger radii in the
constant density CSM model, this light curve has a flatter plateau and
it takes a longer time to rise to peak luminosity.  For this reason,
the light curve plotted in Figure~\ref{fig:model2} has been shifted in
time by $-$40 days due to a larger delay between the time of explosion
and the time of discovery by visual observers.  Although this is a
crude model, the resulting light curves are very similar to light
curves of SN/CSM interaction produced using more detailed 1-D and 2-D
hydrodynamic simulations (e.g., van Marle et al. 2010; Woosley et al.\
2007).

The drop in luminosity at the end of the plateau arises because the
expanding CDS reaches the outer edge of the dense CSM, at which time
CSM interaction ends.  The time at which this occurs is set by the
choice of the outer radius of the CSM, which is chosen to be 120 days
as seen in SNe IIn-P, and the speed at which the CDS moves through the
CSM.  This is easily adjusted and is not a critical part of the model
for the Crab, however, since there are not (yet) any observational
constraints on the duration (or existence) of a plateau in SN~1054.
The steep drop in the model plateau is somewhat artificial, dictated
by the assumed instantaneous drop in density at the outer boundary of
the CSM in the simple model; in reality, it is likely that this
transition would be smoother, as observed in SNe IIn-P.

After the plateau ends and the luminosity drops at $\sim$120 days,
there is a tail of declining luminosity.  There are two likely
contributions to this late-time luminosity.  First, even though the
CDS is no longer sweeping into dense CSM ahead of the shock, there
would still be some luminosity contributed by ongoing shock
interaction as the remaining inner SN ejecta catch up to the dense and
slowly expanding CDS.  This shock luminosity drops with time because
of the shrinking difference between CDS speed (now coasting at
constant speed) and the speed of freely expanding SN ejecta (slower
ejecta take longer to catch up to the CDS).  The strength of this
ongoing shock interaction is set by the final mass and speed of the
CDS at the end of the plateau.  The second expected contribution to
the late-time luminosity is from radioactive decay, which is observed
to be low in SNe IIn-P and should be low in SN~1054 due to the low
abundance of Fe-group elements in the Crab. Even if we allow for some
$\gamma$-ray leakage (see above), the implied masses of $^{56}$Ni
synthesized in the explosion are of order 0.01 $M_{\odot}$.  The solid
orange curves at late times show the sum of the radioactive decay and
shock luminosity, while each individual contribution is shown
separately with dotted and dashed curves.  This late-time luminosity
does not include any contribution from luminosity that might be
associated with early spin-down of the young neutron star;
core-collapse simulations (e.g., Ott et al.\ 2006) yield shorter
periods than one expects from extrapolating the present-day spin
period of 33 ms for the Crab.  Some loss of rotational energy is
probably required, although the mechanism and possible luminosity
associated with this are uncertain.

Both models in Figures~\ref{fig:model1} and \ref{fig:model2} with
somewhat different CSM density distributions provide an adequate
account of the peak luminosity and late time luminosity of SN~1054,
and morover, the final properties of the coasting CDS match the mass,
speed, and kinetic energy of the Crab filaments.  The model with
density falling as $R^{-2}$ would seem to do a better job of
accounting for the fact that SN~1054 was only at its peak luminosity
for a relatively short time of 23 days, whereas the longer plateau
that arises from constant density would seem to match some of the
SN~IIn-P light curves a bit better.  The distinction between these two
is subtle, and proper radiative transfer may provide better
constraints on the emergent temperature and bolometric correction
(these plots are just the total bolometric radiated luminosity).  The
true density distribution may be between the two cases illustrated
here, and it may of course be non-spherical, but the order of
magnitude in the CSM density and mass must be roughly correct.

Although the model used here is quite simplified, it adequately
demonstrates that a relatively low-energy 10$^{50}$ erg explosion that
would be expected from an ecSN can produce both the high peak
luminosity and low late-time luminosity of SNe IIn-P and SN~1054. It
also securely demonstrates that CSM interaction allows the light curve
to be reconciled with the present-day observed properties of the Crab
filaments.  Since the forward shock has been decelerated by the dense
CSM, no additional mass or SN energy needs to be hidden outside the
observed Crab filaments in this model.

\subsection{A Sequence of Events for the Crab}

\begin{figure}\begin{center}
   \includegraphics[width=2.9in]{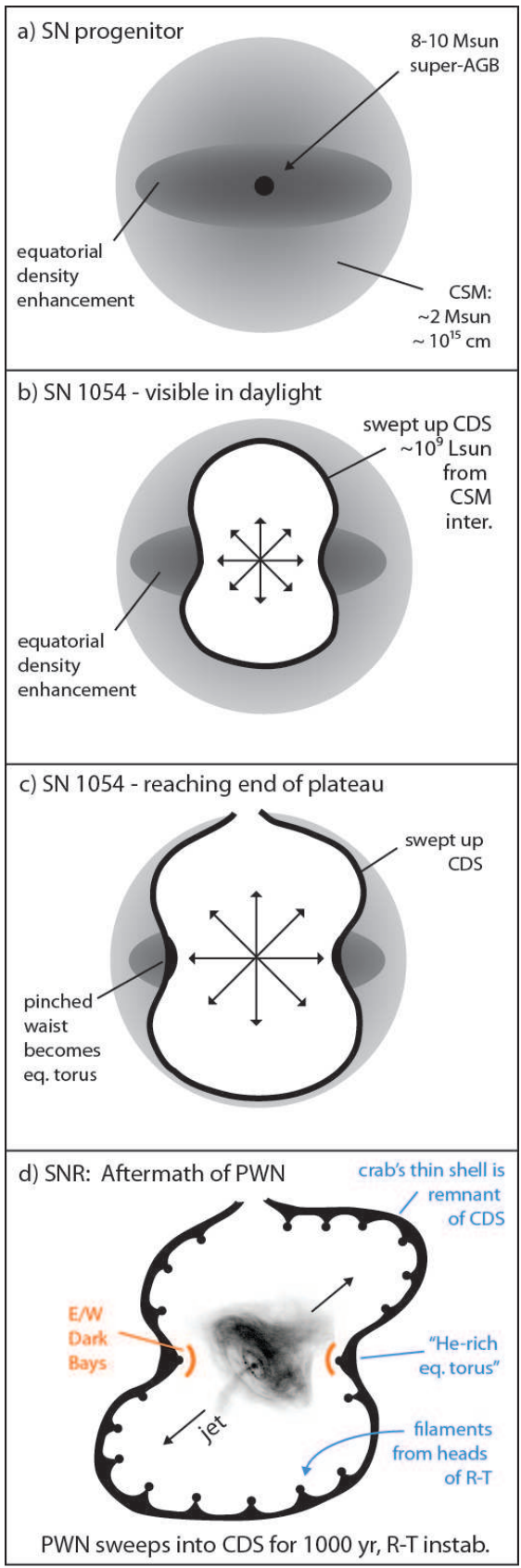}
\end{center}
\caption{Sketch of a possible sequence of events. See text \S 3.4.}
\label{fig:sketch}
\end{figure}

How did the complex structure of the Crab Nebula we see today --
including its spectacularly complex web of dense filaments -- arise as
a result of the CSM interaction model described above?
Figure~\ref{fig:sketch} illustrates a possible sequence of events that
would lead to the basic structures seen in the scenario where SN~1054
was a Type IIn-P explosion.

In this model, the progenitor star must have been a relatively
low-mass (8-10 $M_{\odot}$) super-AGB star surrounded by a dense shell
of CSM within about 1-2 $\times$10$^{15}$ cm of the central star
(Figure~\ref{fig:sketch}a).  To account for some specific structures
seen in the Crab (see below), this initial configuration includes a
density enhancement near the equatorial plane of the progenitor star,
with this ``disk'' seen roughly edge-on and with an east/west
orientation.  This disk is not needed to explain the light curve.

Immediately after core collapse, the remaining stellar envelope that
had not already been shed in the pre-SN mass loss expands outward into
the dense CSM (Figure~\ref{fig:sketch}b).  This collision, which lasts
of order 120 days, produces the peak luminosity phase of SN~1054 as
ejecta kinetic energy is converted to visual-wavelength radiation.  In
this process, the slow CSM is accelerated and the fast SN ejecta are
decelerated; both pile up in the very thin cold dense shell (CDS) that
expands outward at about 1000 km s$^{-1}$.  The shell is thin because
it collapses to a dense layer as radiation cools the gas and removes
pressure support from the post-shock region.

This intense phase of interaction continues until the shock reaches
the outer boundary of the dense CSM (Figure~\ref{fig:sketch}c).  After
that time, the luminosity would plummet (at a decline rate dictated by
the steepness in the density drop, which is not constrained by
available observations).  By this time, about 5 $M_{\odot}$ has piled
up in the slowly expanding CDS, which would coast at a speed of
$\sim$1100 km s$^{-1}$ after this time.  The thin shell is composed
entirely of SN ejecta and CSM ejected before the explosion, and its
chemical abundances reflect the He-rich abundances in the star's
envelope in the final phases of its evolution.  Some of the slower
ejecta continue to crash into the reverse shock and heat the CDS, but
this ongoing shock interaction is far less intense than during the
plateau phase.

After this, the CDS would simply continue to coast, were it not for
the fact that SN~1054 also gave birth to a pulsar than energizes a
PWN.  For the next $\sim$1000 yr up until modern times, the expansion
of the PWN must contend with a massive, slow-moving shell that is in
its path of expansion. It is this last phase of the PWN interacting
with the thin shell that produces much of the complex structure seen
in the Crab (Figure~\ref{fig:sketch}d).  It is known that the
expansion of the PWN has pushed and accelerated the Crab filaments,
since their measured proper motions would seem to indicate an origin
70-80 yr after 1054 A.D. (Trimble 1968; Wyckoff \& Murray 1977;
Bietenholtz et al.\ 1991; Nugent 1998).  In this scenario, where a
thin shell is accelerated outward by an underlying fluid, the thin
shell will be subject to severe Rayliegh-Taylor (RT) instabilities,
and the thin shell will fragment into a network of dense filaments.
As the PWN tries to push through the spaces between these filaments,
it could give rise to the bubble-like morphology in the outer
``[O~{\sc iii}] skin'' seen in deep exposures of the Crab (Fesen \&
Gull 1986).  In some locations, the PWN may push through the shell,
perhaps leading to breakout structures akin to the ``chimney'' or
``jet'' on the northern perimeter of the Crab (Fesen \& Gull 1982; Cox
et al.\ 1991), although this is speculative.  Hydrodynamic simulations
of the PWN pushing into a massive CDS that results from a Type IIn
explosion might be illuminating. 

The overall non-spherical geometry can also be accounted for in this
interaction.  Fesen, Martin, \& Shull (1992) have provided a detailed
discussion of the prominent indentations in the Crab synchrotron
nebula known as the east and west ``bays''.  These features mark a
location where the continuum synchrotron emission is absent,
apparently because the expansion of the PWN has been thwarted there.
Fesen et al.\ (1992) suggested that this may be the result of an
equatorial disk-like distribution of dense CSM around the progenitor
star, that pinched the waist of the expanding remnant, and they also
pointed out a connection to the belt of He-rich filaments that share
the same basic orientation as the E and W bays but at larger radii
(see also Uomoto \& MacAlpine 1987; MacAlpine et al.\ 1989; Lawrence
et al.\ 1995).  This is quite easily accommodated in the CSM
interaction scenario for the Crab if the pre-SN CSM shell had an
equatorial density enhancement, with an equator/pole density contrast
of oder 2.  The common orientation of the bays and the He-rich torus
are harder to accommodate in the ``standard'' view of the Crab, where
the observed Crab filaments are solely the result of the PWN expanding
into the freely expanding SN ejecta (Hester 2008); i.e.  no
interaction with the CSM has occurred here because the hypothetical
forward shock is located far outside the Crab.  In this scenario, the
SE/NW elongation axis of the Crab would arise because of a stronger
push along this axis from the Crab's synchrotron jet, known to share
this orientation (see Fig.~\ref{fig:sketch}d).

\section{PREDICTIONS AND IMPLICATIONS}

\subsection{Light Echoes}

The bright peak luminosity phase of a SN event sends a pulse of light
into its surroundings, which can be seen after a time delay if the
light scatters off nearby dust clouds.  Such ``light echoes'' were
seen in the surroundings of SN~1987A (e.g., Crotts et al.\ 1989;
Sugerman et al.\ 2005), and light echoes have now been detected in
images and studied spectroscopically for several historical SNe (Rest
et al.\ 2005, 2008a; Krause et al.\ 2008b).  Light echoes from Cas A
determined that it was a Type IIb SN (Rest et al.\ 2008b, 2011a,
2011b; Krause et al.\ 2008a), even though it is unclear if its 17th
century SN event was ever observed directly.  Light echoes have even
been detected from the 19th century Great Eruption of $\eta$ Carinae
(Rest et al.\ 2012).  Because $\eta$ Carinae's peak luminosity was
much lower than a SN, and because its peak apparent magnitude was much
fainter than SN~1054, there may yet be hope that continuing efforts
will be able to identify light echoes from SN~1054, despite its old
age of $\sim$1000 yr and low-density environment.  This is a difficult
task, but in the event that light echoes of SN~1054 can be identified,
they offer the opportunity to directly test the model proposed here.
This prediction may provide a rather definitive and useful test.

The ``standard'' picture for the Crab (Hester 2008) places most of the
mass and most of the kinetic energy outside the filaments in a fast,
freely expanding envelope.  At the present time this envelope is
transparent and invisible, and produces no detectable shock as it
encounters the surrounding medium.  During the peak luminosity phase
of SN~1054, however, the fast envelope would dominate the observed
spectrum with a recombination photosphere moving back through the
rapidly expanding ejecta.  In this scenario, the spectrum should show
only broad (5000-10000 km s$^{-1}$) emission lines with similarly
broad P Cygni profiles that should appear to be typical of normal SNe
II-P.

On the other hand, if the CSM interaction model advocated here is
correct, the SN ejecta are quickly decelerated by dense CSM, giving
rise to a high luminosity generated by the collision.  In the CSM
interaction scenario, the peak luminosity spectrum should exhibit
narrow lines typical of SNe IIn.  This might include some broad wings
or intermediate-width components that are typically seen in SNe IIn,
of course, but the strongest emission in the cores of the lines should
be narrow (less than about 1000-1500 km s$^{-1}$, matching the speed
of the CDS).  In particular, the spectra of light echoes from SN~1054
should closely resemble the spectra of the subclass of SNe~IIn-P, like
SNe~1994W, 2009kn, and 2011ht (Chugai et al.\ 2004; Kankare et al.\
2012; Mauerhan et al.\ 2013a), for which high-quality spectra exist.

In addition, the brightness evolution of light echoes can provide
information about the historical light curve, usually providing better
time sampling than the original historical observations. This would be
very useful, given the sparse observations of SN~1054.  The light
curves derived from images of light echoes of SN~1054 may show a
plateau that drops after 120 days if SN~1054 really is a SN IIn-P, or
a smoother decline if it is a more traditional SN IIn (or some other
type of SN).  However, the pleateau is less important than the
presence of a Type~IIn spectrum in order to confirm the essence of the
suggested model, since a normal Type II-P will also have a plateau.

\subsection{A Closer Look at Other Crab-like Remnants}

The scenario suggested here, wherein the Crab is the result of a
Type~IIn SN, may be relevant to the larger class of Crab-like SNRs and
PWNe in general.  The Crab is the archetype for a class of
filled-center supernova remnants (SNRs) known as PWNe (Gaensler \&
Slane 2006), also sometimes referred to as filled-center remnants or
``plerions'' (e.g., Gaensler 2001; Reynolds 1985).  The filled center
refers, of course, to the bright pulsar-powered synchrotron nebula in
the interior of the Crab and its siblings.  The class includes other
famous objects like the Crab's twin SNR~0540-69.3 with its young 50 ms
pulsar (e.g., Morse et al.\ 2006; Seward et al.\ 1984), and 3C58
(Weiler \& Seielstad 1971; Weiler 1983) that is possibly associated
with SN~1181 (Clark \& Stephenson 1977).

It is natural to ask if the physics of a Type~IIn explosion can yield
favorable conditions to enhance the observability or physical
properties of PWNe, or to change their physical properties in a
systematic way.  Namely, SNe~IIn are distinct from other types of SNe
in that the end product of the explosion is a slow, thin, and very
massive (5-10 $M_{\odot}$ or more) dense shell of swept-up CSM and
ejecta.  If a pulsar is born in the core-collapse event associated
with a SN~IIn, its PWN nebula must push outward against this slow and
massive shell.  Essentially, it may be possible that the CDS in a
SN~IIn acts as a ``cage'' to confine the PWN, restricting its
expansion and thereby increasing the energy density inside the volume
of the PWN.  This may make the PWN appear brighter than it otherwise
would be.\footnote{Note that this could be true of SNe IIn-P from
  low-mass progenitors, as well as for other SNe IIn from more massive
  stars (see \S 4.5).}  A pulsar born in a normal SN~II-P would only
need to push outward against the slowest inner ejecta of the SN, which
constitute a much smaller fraction of the total mass of the progenitor
star, and may not as easily confine the PWN.

\subsection{The Invisible Fast Envelope and Forward Shock of
  the Crab}

If the Crab Nebula is the end product of a Type~IIn SN explosion, as
described above, then there is no need to invoke the existence of a
hypothetical extended, rapidly expanding SN envelope outside the Crab
that retains most of the mass and energy of the exploded star (Hester
2008).  In a model where a Type~IIn event powers the high luminosity
of SN~1054 despite the low explosion energy, the outer edge of the
Crab's shell of thermal filaments is the remnant of the CDS.  The lack
of any observational evidence for the putative fast envelope and the
lack of any observed signature of a blast wave far outside the Crab
are both, therefore, expected if SN~1054 was a Type IIn event.  This
model predicts that continued efforts to measure a massive outer fast
envelope will only produce deeper upper limits.  There is a
possibility that there could be a remaining exterior blast wave if the
forward shock accelerated through a steep density gradient after
exiting a dense CSM shell (see Smith 2013), but this would involve
very little mass and kinetic energy, and has probably cooled
significantly in the past 1000 yr since explosion.  We do not expect
to see an extended, fast, cold neutral ejected stellar envelope with
any substantial mass (several $M_{\odot}$).  If there are any
surrounding clouds that show evidence for some outward acceleration by
radiation pressure from the SN, it is likely that their accelerated
mass and momentum are far less than would be expected for a 10$^{51}$
erg blast wave.

\subsection{Post-Shock Dust Formation}

Recent observations have placed much better constriants on the mass of
dust that resides in the dense filaments of the Crab Nebula.  Mid-IR
observations with {\it Spitzer} (Temim et al.\ 2006, 2012) and far-IR
observations with {\it Herschel} (Gomez et al.\ 2012) yield a
surprisingly large mass of 0.1--0.2 $M_{\odot}$ of dust.  With a
filament gas mass of order 5 $M_{\odot}$, this would suggest a
gas/dust mass ratio of 25-50, significantly higher than in the normal
insterstellar medium, and indicative of very efficient dust
condensation. This indicates that the dust formed from processed
stellar ejecta.  This dust mass is much higher than that which is
deduced from IR measurements of normal SNe II-P, typically of order
10$^{-3}$ $M_{\odot}$ or less (Sugerman et al.\ 2006; Andrews et al.\
2010; Meikle et al.\ 2007).  The IR images of the Crab have sufficient
angular resolution to suggest that most of the emitting dust resides
in the filaments, and not in a hypothetical envelope outside the Crab.
This confirms the indication that dust resides in the filaments
themselves, based on the fact that some filaments on the near side of
the Crab's shell are seen in silhouette against the synchrotron
continuum (Fesen \& Blair 1990).

If the Crab is the result of a Type~IIn supernova, then CSM
interaction that formed the thin CDS might also have precipitated
efficient dust formation.  Studies of the Type~Ibn event SN~2006jc
showed that dust formed efficiently and quickly ($\sim$50 days after
explosion) as the SN ejecta crashed into dense He-rich CSM (Smith et
al.\ 2008a).  Mounting evidence from studies of SNe~IIn suggest that
these compressed post-shock layers may be efficient sites of dust
formation in H-rich SNe~IIn as well (see Smith et al.\ 2012, and
references therein). The efficient raditaive cooling and very high
densities in the post-shock zones may lead to much more efficient and
rapid dust formation in SNe IIn as compared to normal SNe where the
ejecta expand rapidly and the density quickly drops.  The drop in
luminosity after the plateau of SNe~IIn in particular may aid the
formation of dust, because the CDS temperature will drop while the
density is still very high.  Indeed, the very low late-time luminosity
and red color of SNe IIn-P like SN~2011ht seem to suggest efficient
dust formation (Mauerhan et al.\ 2013a).  Such rapid dust formation in
the filaments of the Crab is harder to explain in the standard model
of SN~1054 as a normal SN, which lacks such a dense and thin cooling
layer.

\subsection{The Diverse Progenitors of SNe IIn}

Taken together, the Crab Nebula and the sub-class of SNe IIn-P provide
strong evidence that a subset of SNe~IIn arise from the lowest-mass
progenitor stars that can undergo core collapse.  These are super-AGB
stars in the range 8-10 $M_{\odot}$ that suffer an ecSN rather than an
Fe core collapse event.

In that case, the class of SNe~IIn must be a fairly heterogeneous
collection of explosions, with both very high mass
progenitors and low-mass progenitors.  In addition to the low-mass
8-10 $M_{\odot}$ super-AGB stars that may produce SNe~IIn-P, some
SNe~IIn may arise from star systems with even lower initial masses
below 8~$M_{\odot}$; these are the so-called ``hybrid'' Type Ia/IIn
objects, that seem to result from Type Ia explosions surrounded by
dense H-rich CSM, such as SN~2002ic (e.g., Chugai \& Yungelson 2004;
Silverman et al.\ 2013; and references therein).

At the other extreme, there are several arguments that SNe~IIn are
associated with very massive progenitor stars as well. Very luminous
SNe~IIn like SN~2006gy, SN~2006tf, and others require CSM masses of
order 10-20 $M_{\odot}$ in order to power their high luminosity with
CSM interaction (Smith et al.\ 2007, 2008b, 2010a; Smith \& McCray
2007; van Marle et al.\ 2010; Woosley et al.\ 2007), so their
progenitors must have been very massive stars.  Even moderately
luminous SNe IIn seem to require mass-loss rates that can only be
achieved with the eruptive modes of mass loss seen in LBVs (Gal-Yam et
al.\ 2007; Smith et al.\ 2007, 2008b).  There have been 3 clear
detections of hypergiant LBV-like progenitors of SNe~IIn, including
SN~2005gl (Gal-Yam \& Leonard 2009; Gal-Yam et al.\ 2007), SN~1961V
(Smith et al.\ 2011b; Kochanek et al.\ 2011), and most recently
SN~2009ip (Mauerhan et al.\ 2013b; Smith et al.\ 2010; Foley et al.\
2011).  The case of SN~2009ip was particularly interesting, seen as an
eruptive LBV that was studied spectroscopically and photometrically
before it exploded.  Additionally, a luminous blue source is seen at
the location of the luminous SN~IIn 2010jl, although the SN has not
yet faded; this source is either an extremely luminous and massive
supergiant or a very young star cluster.  Either option would require
an initial mass above $\sim$30 $M_{\odot}$ (Smith et al.\ 2011c).
Thus, there is strong and direct evidence that very massive LBV-like
stars do sometimes explode as SNe~IIn.  Some lower-luminosity SNe~IIn
may arise from intermediate masses too, such as 20-40 $M_{\odot}$ red
supergiants with extreme winds (see Smith et al.\ 2009a, 2009b).

Thus, the diverse progenitors of various SNe~IIn span a wide range of
initial masses.  This must be taken into account when interpeting
results of the statistical distribution of SNe~IIn in galaxies, as
compared to other types of SNe (e.g., Anderson et al.\ 2012; Kelly \&
Kirshner 2010).

\subsection{SN Impostors and ecSNe}

The model discussed above casts SN~1054 and the Crab Nebula as the
result of a Type IIn-P explosion, similar to SNe~1994W, 2009kn, and
2011ht, where the explosion mechanism was the collapse of a degenerate
ONeMg core (i.e. an ecSN) that yields a sub-energetic (10$^{50}$ erg)
explosion and low $^{56}$Ni mass.  A key component of the proposed
model is that intense CSM interaction permits the resulting SN to be
more luminous than a traditional 10$^{51}$ erg core-collapse SN, but
with an order of magnitude lower explosion energy.  This reconciles
the apparent brightness of SN~1054 with the abundances and low kinetic
energy in the Crab.

There are two other classes of explosions that are commonly discussed
as possible ecSNe as well.  One is the class of sub-luminous SNe II-P
that make up the bottom end of the luminosity distribution for SNe
II-P (Pastorello et al.\ 2004, 2009), and which also have directly
detected progenitor stars that are near the low-mass end of the
distribution of SN II-P progenitors (Smartt 2009).  The low-luminosity
SNe II-P show no clear signs of CSM interaction in their spectra, with
broader emission lines and P Cygni absorption features that are unlike
LBVs or SNe~IIn (Smith et al.\ 2009c).  In that case, there is no
apparent conflict with the ecSN model suggested here for the Crab,
since at least in principle, an ecSN may or may not have dense-enough
CSM to raise its luminosity appreciably.  With the expected explosion
energy and $^{56}$Ni yield, the luminosities of the faintest SNe II-P
match expectations for ecSN with no significant CSM interaction.

Another class of transients that has been linked to ecSNe by some
authors (Thompson et al.\ 2009; Botticella et al.\ 2009) is the
sub-class of SN impostors whose archetypes are SN~2008S and the 2008
transient in NGC~300 (NGC~300-OT).  The motivations for this link are
that (1) the progenitors are consistent with short-lived
dust-enshrounded stars of relatively low initial mass (Prieto et al.\
2008, 2009; Thompson et al.\ 2009), reminiscent of the super-AGB stars
expected as the progenitors of ecSNe, and (2) various factors suggest
that the transients are explosive (e.g., Kochanek 2011).  With
durations of $\sim$100 d and peak absolute magnitudes around $-$14
mag, the total radiated energy of these events is a few 10$^{47}$ erg,
or a fraction of only $\sim$10$^{-3}$ of the expected explosion energy
of an ecSN.  This is a factor of 10 lower than the luminosity and
radiated energy in the low-luminosity SNe~II-P like SN~2005cs.  

A potential conflict in this picture is that the spectra of these
transients exhibit narrow lines and no significant broad absorption at
the times of peak luminosity, similar to the spectra of SNe~IIn.
This, in turn, suggests that they are dominated by strong CSM
interaction.  Yet, these SN impostor transients are a factor of
$\sim$5 less luminous than even the faintest SNe II-P discussed above.
CSM interaction can generate a lower luminosity if the dense CSM only
occupies a small fraction of the solid angle of the explosion, but in
that case the spectrum should also reveal the bare ecSN photosphere
(which should show broad lines similar to SN~2005cs), and moreover,
the level of dust obscuration around the progenitors seems to require
that the dense CSM covers a large fraction of the solid angle. The CSM
mass of order 1 $M_{\odot}$ around these progenitors (Wesson et al.\
2010) and the observed line widths would require that CSM interaction
from a 10$^{50}$ erg explosion would either be several times more
luminous or would last much longer than 100 days (i.e. as in the 10 yr
long Great Eruption of $\eta$~Carinae, if it was powered by CSM
interaction in a 10$^{50}$ erg explosion; Smith 2013), and would
convert a larger fraction of the kinetic energy into light.  (It is
not clear yet if the integrated IR luminosity can make up the
difference, but perhaps continued study will answer this.)  Because an
ecSN that interacts with CSM should be significantly more luminous
than the faint SNe~II-P, not less luminous, this suggests that SN
impostors probably arise from explosive transients of lower energy
(10$^{48}-10^{49}$ ergs).  Unless an ecSN can produce an explosion
energy this low, there is probably some other physical mechanism for
these SN impostors and related transients that is different from the
Crab and SNe IIn-P.  Discussing these numerous other possibilities is
beyond the scope of this paper.

\smallskip\smallskip\smallskip\smallskip
\noindent {\bf ACKNOWLEDGMENTS}
\smallskip
\footnotesize

I thank the anonymous referee for helpful comments, and I thank Rob
Fesen for interesting discussions about the observed properties of the
Crab Nebula.



\begin{thebibliography}\scriptsize

\bibitem[]{} Arnett, W.D.\ 1974, ApJ, 193, 169

\bibitem[]{} Anderson, J., et al.\ 2012, MNRAS, 424, 1372

\bibitem[]{} Andrews, J.E., et al.\ 2010, ApJ, 715, 541

\bibitem[]{} Arcavi, I., et al.\ 2012, ApJ, 756, L30

\bibitem[]{} Bietenholz, M.F., Kronberg, P.P., Hogg, D.E., \& Wilson,
  A.S.\ 1991, ApJ, 373, L59

\bibitem[]{} Blair, W.D., et al.\ 1997, ApJS, 109, 473

\bibitem[]{} Botticella, M.T>, et al.\ 2009, MNRAS, 398, 1041

\bibitem[]{} Brecher, K., Fesen, R.A., Maran, S.P., \& Brandt, J.C.\
  1983, Obs., 103, 106

\bibitem[]{} Buta, R.J. 1982, PASP, 94, 578 

\bibitem[]{} Chevalier, R.A.\ 1977, in Supernovae, ed. D.\ Schramm
  (Dordrecht: Reidel), 53

\bibitem[]{} Chevalier, R.A., \& Irwin, C.M.\ 2011, ApJ, 729, L6

\bibitem[]{} Chugai, N.N., \& Danziger, I. J.\ 1994, MNRAS, 268, 173

\bibitem[]{} Chugai, N.N., \& Yungelson, L.R.\ 2004, Astronomy Letters, 30, 65

\bibitem[]{} Chugai, N.N., et al.\ 2004, MNRAS, 352, 1213

\bibitem[]{} Clark, D.H., \& Stephenson, R.R.\ 1977, The Historical
  Supernovae (Oxford: Pergamon)

\bibitem[]{} Cox, C.I., Gull, S.F., \& Green, D.A.\ 1991, MNRAS, 250, 750

\bibitem[]{} Crotts, A.P.S., et al.\ 1989, ApJ, 347, L61

\bibitem[]{} Davidson, K., \& Fesen, R.A.\ 1985, ARA\&A, 23, 119

\bibitem[]{} Davidson, K., et al.\ 1982, ApJ, 253, 696

\bibitem[]{} Dessart, L., Hillier, D.J., Li, C., \& Woosley, S.\ 2012,
  MNRAS, 424, 2139

\bibitem[]{} Duyvendak, J.J.L.\ 1942, PASP, 54, 941

\bibitem[]{} Elias-Rosa, N., et al.\ 2009, ApJ, 706, 1174

\bibitem[]{} Falk, S.W., \& Arnett, W.D.\ 1977, ApJS, 33, 515

\bibitem[]{} Fesen, R.A., \& Gull, T.R.\ 1986, ApJ, 306, 259

\bibitem[]{} Fesen, R.A., \& Blair, W.D.\ 1990, ApJ, 351, L45

\bibitem[]{} Fesen, R.A., \& Kirshner, R.P.\ 1982, ApJ, 258, 1

\bibitem[]{} Fesen, R.A., Martin, C.L., \& Shull, J.M.\ 1992, ApJ, 399, 599

\bibitem[]{} Fesen, R.A., Shull, J.M., \& Hurford, A.P.\ 1997, AJ,
  113, 354

\bibitem[]{} Frail, D.A., Kassim, N.E., Cornwell, T.J., \& Goss, W.M.\
  1995, ApJ, 454, L129

\bibitem[]{} Foley, R.J., et al.\ 2011, ApJ, 732, 32

\bibitem[]{} Gal-Yam, A., \& Leonard, D.C.\ 2009, Nature, 458, 865

\bibitem[]{} Gal-Yam, A., et al.\ 2007, ApJ, 656, 372

\bibitem[]{} Gaensler, B.M.\ 2001, in Young Supernova Remnants (New
  York: Melville), 295

\bibitem[]{} Gaensler, B.M., \& Slane, P.O.\ 2006, ARA\&A, 44, 17

\bibitem[]{} Glaccum, W., et al.\ 1982, BAAS, 14, 612

\bibitem[]{} Gomez, H., et al.\ 2012, ApJ, 760, 96

\bibitem[]{} Graham, J.R., Wright, G.S., \& Longmore, A.J.\ 1990, ApJ,
  352, 172


\bibitem[]{} Hachinger, S., et al.\ 2012, MNRAS, 422, 70

\bibitem[]{} Henry, R.B.C., \& MacAlpine, G.M.\ 1982, ApJ, 258, 11

\bibitem[]{} Hester, J.J.\ 2008, ARA\&A, 46, 127

\bibitem[]{} Hillebrandt, W.\ 1982, A\&A, 110, 3

\bibitem[]{} Kankare, E., et al.\ 2012, MNRAS, 424, 855

\bibitem[]{} Kelly, P.L., \& Kirshner, R.P.\ 2012, ApJ, 759, 107

\bibitem[]{} Kitaura, F.S., Janka, H.T., \& Hillebrandt, W.\ 2006,
  A\&A, 450, 345

\bibitem[]{} Kochanek, C.\ 2011, ApJ, 741, 37

\bibitem[]{} Kochanek, C., et al.\ 2011, ApJ, 737, 76

\bibitem[]{} Kirshner, R.P.\ 1974, ApJ, 194, 323

\bibitem[]{} Krause, O., et al.\ 2008a, Science, 320, 1195

\bibitem[]{} Krause, O., et al.\ 2008b, Nature, 456, 617

\bibitem[]{} Lau, H.H.B., Gil-Pons, P., Doherty, C., \& Lattanzio, J.\
  2012, A\&A, 542, A1

\bibitem[]{} Lawrence, et al.\ 1995, AJ, 109, 2635

\bibitem[]{} Leonard, D. C., et al.\ 2002, PASP, 114, 35

\bibitem[]{} Li, W., et al.\ 2011, MNRAS, 412, 1441

\bibitem[]{} Loh, E.D., Baldwin, J.A., \& Ferland, G.J.\ 2010, ApJ, 716, L9

\bibitem[]{} Lundmark, K.\ 1921, PASP, 33, 225

\bibitem[]{} Lundqvist, P., Fransson, C., \& Chevalier, R.A.\ 1986,
  A\&A, 162, L6

\bibitem[]{} MacAlpine, G.M., McGaugh, S.S., Mazzarella, J.M., \&
  Uomoto, A.\ 1989, ApJ, 342, 364

\bibitem[]{} MacAlpine, G.M., \& Satterfield, T.J.\ 2008, AJ, 136,
  2152

\bibitem[]{} Marsden, P.L., et al.\ 1984, ApJ, 278, L29

\bibitem[]{} Mayall, N.U., \& Oort, J.H.\ 1942, PASP, 54, 95

\bibitem[]{} Mauche, C.W., \& Gorenstein, P.\ 1989, ApJ, 336, 843

\bibitem[]{} Mauerhan, J.C., Smith, N., Silverman, J.M., Filippenko,
  A.V., Morgan, A.N., Cenko, S.B., Ganeshalingam, M., Clubb, K.I.,
  Matheson, T.\ 2013a, MNRAS, 431, 2599 

\bibitem[]{} Mauerhan, J.C., et al. 2013b, MNRAS, 430, 1801 

\bibitem[]{} Meikle, W.P.S., et al.\ 2007, ApJ, 665, 608

\bibitem[]{} Miller, J.S.\ 1973, ApJ, 180, L83

\bibitem[]{} Minkowski, R.\ 1970, PASP, 82, 470

\bibitem[]{} Minkowski, R.\ 1971, IAUS, 46, 241

\bibitem[]{} Miyajo, S., Nomoto, K., Yokoi, K., \& Sugimoto, S.\ 1980,
  PASJ, 32, 303

\bibitem[]{} Morse, J.A., Smith, N., Blair, W.P., Kirshner, R.P.,
  Winkler, P.F., \& Hughes, J.P.\ 2006, ApJ, 644, 188

\bibitem[]{} Nomoto, K., Sugimoto, D., Sparks, W.M., Fesen, R.A.,
  Gull, T.R., \& Miyaji, S.\ 1982, Nature, 299, 803

\bibitem[]{} Nomoto, K.\ 1987, ApJ, 277, 791

\bibitem[]{} Nugent, R.L.\ 1998, PASP, 110, 831

\bibitem[]{} Ott, C.D., Burrows, A., Thompson, T.A., Livne, E., \&
  Walder, R.\ 2006, ApJS, 164, 130

\bibitem[]{} Pastorello, A., et al.\ 2007, Nature, 474, 892

\bibitem[]{} Pastorello, A., et al.\ 2009, MNRAS, 394, 2266

\bibitem[]{} Pequignot, D., \& Dennefeld, M.\ 1983, A\&A, 120, 249

\bibitem[]{} Poelarands, A.J.T., Herwig, F., Langer, N., \& Heger, A.\
  2008, ApJ, 675, 614

\bibitem[]{} Predehl, P., \& Schmitt, J.H.M.M.\ 1995, A\&A, 293, 889

\bibitem[]{} Prieto, J.L., et al.\ 2008, ApJ, 681, L9

\bibitem[]{} Prieto, J.L., et al.\ 2009, ApJ, 705, 1425

\bibitem[]{} Rest, A., et al.\ 2005, Nature, 438, 1132

\bibitem[]{} Rest, A., et al.\ 2008a, ApJ, 680, 1137

\bibitem[]{} Rest, A., et al.\ 2008b, ApJL, 681, L81

\bibitem[]{} Rest, A., et al.\ 2011a, ApJ, 732, 3

\bibitem[]{} Rest, A., et al.\ 2011b, ApJ, 732, 2

\bibitem[]{} Rest, A., et al.\ 2012, Nature, 482, 375

\bibitem[]{} Reynolds, S.\ 1985, in The Crab Nebula and Related
  Supernova Remnants (Cambridge: Cambridge University Press), 173

\bibitem[]{} Rudie, G.C., Fesen, R.A., \& Yamada, T.\ 2007, MNRAS,
  384, 1200

\bibitem[]{} Sankrit, R., et al.\ 1998, ApJ, 504, 344

\bibitem[]{} Satterfield, T.J., Katz, A.M., Sibley, A.R., MacAlpine,
  G.M., \& Uomoto, A.\ 2012, AJ, 144, 27

\bibitem[]{} Seward, F.D., Harnden, D., \& Helfand, D.\ 1984\ ApJ,
  287, L19

\bibitem[]{} Seward, F.D., Gorenstein, P., \& Smith, R.K.\ 2006, ApJ,
  636, 873

\bibitem[]{} Silverman, J., et al.\ 2013, preprint (arXiv:1304.0763)

\bibitem[]{} Smartt, S.J.\ 2009, ARA\&A, 47, 63

\bibitem[]{} Smith N.\ 2003, MNRAS, 346, 885

\bibitem[]{} Smith N.\ 2013, MNRAS, 429, 2366

\bibitem[]{} Smith, N., \& McCray, R.\ 2007, ApJ, 671, L17

\bibitem[]{} Smith, N., et al.\ 2007, ApJ, 666, 1116 

\bibitem[]{} Smith, N., Foley, R.J., \& Filippenko, A.V.\ 2008a, ApJ,
  680, 568

\bibitem[]{} Smith, N., et al.\ 2008b, ApJ, 686, 467 

\bibitem[]{} Smith, N., Hinkle, K., \& Ryde, N.\ 2009a, AJ, 137, 3558

\bibitem[]{} Smith, N., et al.\ 2009b, ApJ, 695, 1334

\bibitem[]{} Smith, N., et al.\ 2009c, ApJ, 697, L49

\bibitem[]{} Smith, N., Chornock, R., Silverman, J.M., Filippenko,
  A.V., \& Foley, R.J.\ 2010a, ApJ, 709, 856 

\bibitem[]{} Smith, N., et al.\ 2010b, AJ, 139, 1451

\bibitem[]{} Smith, N., Li, W., Filippenko, A.V., \& Chornock, R.\
  2011a, MNRAS, 412, 1522

\bibitem[]{} Smith, N., et al.\ 2011b, MNRAS, 415, 773

\bibitem[]{} Smith, N., et al.\ 2011c, ApJ, 732, 63

\bibitem[]{} Smith, N., et al.\ 2012, AJ, 143, 17

\bibitem[]{} Sollerman, J., et al.\ 2000, ApJ, 537, 861

\bibitem[]{} Sollerman, J., Kozma, C., \& Lundqvist, P.\ 2001, A\&A, 366, 197

\bibitem[]{} Sugerman, B.E.K., et al.\ 2005, ApJS, 159, 60

\bibitem[]{} Sugerman, B.E.K., et al.\ 2006, Science, 313, 196

\bibitem[]{} Sugerman, B.E.K., Andrews, J.E., Barlow, M.J., et al.\
  2012, ApJ, 749, 170

\bibitem[]{} Temim, T., et al.\ 2006, ApJ, 132, 1610

\bibitem[]{} Temim, T., et al.\ 2012, ApJ, 753, 72

\bibitem[]{} Temim, T., \& Dwek, E.\ 2013, preprint (arXiv:1302.5452)

\bibitem[]{} Thompson, T.A., et al.\ 2009, ApJ, 705, 1364

\bibitem[]{} Trimble, V.\ 1968, AJ, 73, 535

\bibitem[]{} Trimble, V.\ 1973, PASP, 85, 579

\bibitem[]{} Uomoto, A., \& MacAlpine, G.M.\ 1987, AJ, 93, 1151

\bibitem[]{} van Marle, A.J., Smith, N., Owocki, S.P., \& van Veelen,
  B.\ 2010, MNRAS, 407, 2305

\bibitem[]{} Wanajo, S., Nomoto, K., Janka, H.T., Kitaura, F.S., \&
  Muller, B.\ 2009, ApJ, 695, 208

\bibitem[]{} Weaver T.A., \& Woosley, S.E.\ 1979, BAAS, 11, 724

\bibitem[]{} Weiler, K.W.\ 1983, Observatory, 103, 85

\bibitem[]{} Weiler, K.W., \& Seielstad, G.A.\ 1971, ApJ, 163, 455

\bibitem[]{} Wesson, R., et al.\ 201, MNRAS, 403, 474

\bibitem[]{} Woltjer, L.\ 1958, Bull.\ Astron.\ Inst.\ Neth., 14, 39

\bibitem[]{} Wood, P.R., \& Faulkner, D.J.\ 1986, ApJ, 307, 659

\bibitem[]{} Woosley et al. 2007, Nature, 450, 390

\bibitem[]{} Wyckoff, S., \& Murray, C.A.\ 1977, MNRAS, 180, 717

\end{thebibliography}
\end{document}